\documentclass[traditabstract, letter]{aa}  
  \usepackage[fleqn]{amsmath}
\usepackage[english]{babel}
\usepackage[utf8]{inputenc}
\usepackage{amsmath,amsfonts,amssymb,enumerate}
\usepackage{txfonts,graphics,graphicx,epstopdf,float,lscape,longtable,dcolumn,footnote,subcaption,caption,color,url,hyperref,listings}
\usepackage{rotating,afterpage}
\usepackage{natbib}
\bibpunct{(}{)}{;}{a}{}{,}	
\newcommand{\rhopdf}{$\rho$-PDF}
\newcommand{\sigmas}{$\sigma_s$}

\newcommand{\Ms}{$\cal{M}_\mathrm{s}$}
\newcommand{\jk}[1]{\textbf{\textcolor{black}{#1}}}
\begin{document} 
   \title{Relationship between turbulence energy and density variance in the Solar neighbourhood molecular clouds}
      \author{J. Kainulainen\inst{1,2}	\and
           C. Federrath\inst{3}
          }

%   \offprints{jtkainul@mpia.de}
	\institute{Dept. of Space, Earth and Environment, Chalmers University of Technology, Onsala Space Observatory, 439 92 Onsala, Sweden \\
              	\email{jouni.kainulainen@chalmers.se}     
				\and 
				Max-Planck-Institute for Astronomy, K\"onigstuhl 17, 69117 Heidelberg, Germany
              	\and    
	     		Research School of Astronomy and Astrophysics, Australian National University, Canberra, ACT 2611, Australia \\
		                    }
   \date{Received ; accepted }
  \abstract
{The relationship between turbulence energy and gas density variance is a fundamental prediction for turbulence-dominated media and it is commonly used in analytic models of star formation. We determine this relationship for 15 molecular clouds in the Solar neighbourhood. We use the linewidths of the CO molecule as the probe of the turbulence energy (sonic Mach number, \Ms ) and three-dimensional models to reconstruct the density probability distribution function (\rhopdf) of the clouds, derived using near-infrared extinction and \emph{Herschel} dust emission data, as the probe of the density variance (\sigmas). 
We find no significant correlation between \Ms\ and \sigmas\ among the studied clouds, however, we also cannot rule out a weak correlation. 
In the context of turbulence-dominated gas, the range of the \Ms\ and \sigmas\ values corresponds with the model predictions. The data cannot constrain whether or not the turbulence driving parameter, $b$, and/or thermal-to-magnetic pressure ratio, $\beta$, vary among the sample clouds. Most clouds are not in agreement with field strengths stronger than given by $\beta \lesssim 0.05$. A model with $b^2 \beta / (\beta+1) = 0.30 \pm 0.06$ provides an adequate fit to the cloud sample as a whole.
When considering the average behaviour of the sample, we can rule out three regimes: (i) strong compression combined with a weak magnetic field ($b \gtrsim 0.7$ and $\beta \gtrsim 3$), (ii) weak compression ($b \lesssim 0.35$), and (iii) strong magnetic field ($\beta \lesssim 0.1$). 
Including independent magnetic field strength estimates to the analysis, the data rule out solenoidal driving ($b < 0.4$) for the majority of the Solar neighbourhood clouds. However, most clouds have $b$ parameters larger than unity, which indicates a discrepancy with the turbulence-dominated picture; we discuss the possible reasons for this.  
}
  \titlerunning{Turbulence energy and density variance in the Solar neighbourhood clouds} 
  \authorrunning{J. Kainulainen \& C. Federrath } 
  \maketitle

%________________________________________________________________

%*************************
%*************************
\section{Introduction} %*
%*************************
%*************************
\label{sec:intro}

% The first paragraph

In the prevalent paradigm of the turbulence-regulated interstellar medium (ISM) and star formation, the star formation rate of the ISM is linked to the internal density distribution of molecular clouds. In this picture, the cloud structure is strongly affected by supersonic turbulent motions that drive the formation of density enhancements, some of which become self-gravitating and form stars \citep[for reviews, see][]{elmegreen_scalo2004, hennebelle_falgarone2012, padoan2014ppvi}. Specifically, theoretical works predict, and star formation rate models assume, that the initial density distribution of the ISM is characterised by a log-normal probability density function of gas densities (hereafter \rhopdf ), a form that naturally results in isothermal, non-gravitating, supersonically turbulent gas \citep[e.g.,][]{vazquez1994, padoan1997nj, passot1998}. The width of the lognormal \rhopdf, $\sigma_\mathrm{s}$, where $s = \ln{\rho/\rho_\mathrm{0}}$ and $\rho_{0}$ is the mean density, depends on turbulence energy, magnetic field strength, and the fraction of compressive energy in the gas \citep{molina2012,federrath_banerjee2015}
\begin{equation}
\sigma_\mathrm{s}^2 = \ln{  \bigg( 1+b^2{\cal M}_\mathrm{s}^2 \frac{\beta}{1+\beta}  \bigg)  },
\label{eq:b} 
\end{equation}
where \Ms\ is the sonic Mach number, $\beta$ is the ratio between thermal and magnetic energies, and $b$ describes the mixture of solenoidal (divergence-free) and compressive (curl-free) modes in the acceleration field that drives the turbulence \citep{federrath2008ks,federrath2010rk}. Eq.~\ref{eq:b} represents a fundamental prediction of the turbulence-regulated ISM paradigm, and it plays a crucial role in analytic theories of star formation by coupling the density statistics of the gas to the physical processes acting in it \citep[e.g.,][for a review, see \citealt{padoan2014ppvi}]{krumholz_mckee2005, padoan_nordlund2011, hennebelle_chabrier2011, federrath_klessen2012}.

% The problematic
However, few works have obtained observational constraints for the relationship given by Eq.~\ref{eq:b}. This is because of the observational difficulties in probing it: none of its parameters are directly accessible to observations. The sonic Mach number can be estimated through measurements of the molecular line widths, however, it is not clear what fraction of the observed line width is caused by turbulence. The \rhopdf\ width cannot be directly measured, because observations only probe the projected column densities, not volume densities. Also, observational techniques all probe only a limited range of column densities \citep[e.g.,][]{goodman2009ps}, making a reliable quantification difficult. It is also difficult to measure magnetic field strengths in clouds, leaving the $\beta$ parameter uncertain. Consequently, the relationship given by Eq.~\ref{eq:b} remains not well constrained.

Works targeting single molecular clouds in the Solar neighbourhood have suggested values of $b\sim 0.5$ \citep[][]{padoan1997nj, brunt2010fp_pdf}. \citet{ginsburg2013} found that turbulence in the massive star formation region W49 is likely driven compressively with $b > 0.4$. Using small samples of clouds to probe the mean value of $b$, \citet{kainulainen_tan2013} derived $b=0.2^{+0.37}_{-0.22}$, assuming non-magnetised gas. \citet{kainulainen2013fh} compared the dense gas mass fractions of molecular clouds with simulations that spanned a variety of parameters \{$b, \beta$,\Ms \} and concluded that only simulations with $b \approx 1/3-0.4$ match the observations of the Solar neighbourhood clouds. \citet{kainulainen2013fh} also found that infrared dark clouds showed possibly larger $b$ values. The above works tend to suggest dominance of compressive driving modes ($b\gtrsim 0.4$) in the clouds located in the Galactic disc and spiral arms. However, \citet{federrath2016brick} found $b = 0.22 \pm 0.12$ in the central molecular zone cloud G0.253+0.016 (‘The Brick’), which indicates that the driving is primarily solenoidal in environments that are dominated by strong shearing motions. In summary, the observational works constraining $b$ are still few and based on individual measurements or small samples, and they employ different approaches to estimate $b$; systematic studies are yet missing.

Meanwhile, numerical simulations of molecular cloud formation in disc galaxies and idealized colliding flow setups predict a range of $b$ values emerging in different cloud environments and at different times \citep{jin2017,koertgen2017driving}. These studies suggest that a single constant $b$ value cannot be used to describe the turbulence driving of all clouds, but instead that $b$ can vary significantly from cloud to cloud, covering the full range from solenoidal to compressive driving. 

% What we bring to the table: rho-PDFs. 

Enabling progress in the topic, we have developed a technique to determine the volumetric PDF widths (\sigmas) from projected (column) density maps. We have previously used the technique to perform the first systematic observational quantification of the PDF widths in the Solar neighbourhood  clouds \citep{kainulainen2014}. In this paper, we analyse the \Ms -- $\sigma_\mathrm{s}$ relationship in the Solar neighbourhood clouds using the \sigmas\ data from \citet{kainulainen2014} and \Ms\ data from literature. We also derive new \sigmas\ values for the Orion A, Orion B, and California molecular clouds. This allows us to confront the Eq.~\ref{eq:b} with the latest observational results and to provide constraints for the analytic star formation rate models using Eq.~\ref{eq:b}

%*******************************************
%*******************************************
\section{Data} %*
%*******************************************
%*******************************************
\label{sec:data}

% Intro: values from literature.

We adopt most data for \sigmas\ and \Ms\ from literature. \citet{kainulainen2014} derived \sigmas\ values for several Solar neighbourhood clouds and we use these values as such (listed in Table \ref{tab:parameters}). In short, their technique was based on reconstructing the observed column density data with the help of an assemblage of three-dimensional model forms (prolate spheroids). The model forms are arranged hierarchically, nested inside each others, which allows construction of complex structures, such as fractals, containing filaments and sheets, similar to observed cloud structures. The technique was used in conjunction with near-infrared dust extinction derived column density maps from \citet{kainulainen2009probing} that had the spatial resolution of 0.1 pc and they probed the column densities between $N$(H$_2$)$\approx 1-25 \times 10^{21}$ cm$^{-2}$. \citet{kainulainen2014} estimated the uncertainty of the resulting \sigmas\ values to be about 20\%. We refer to \citet{kainulainen2014} for further details.

We adopt \Ms\ values for most clouds from \citet{kainulainen_tan2013} who used CO (1-0) line emission data from \citet{dame2001} to estimate the total linewidths of the clouds (listed in Table \ref{tab:parameters}). The sonic Mach numbers are computed from the linewidths using
\begin{equation}
{\cal M}_\mathrm{s} = \frac{\sqrt{3}\sigma_\mathrm{v}^{1\mathrm{D}}}{c_\mathrm{s}},
\label{eq:ms}
\end{equation}
where $\sigma_\mathrm{v}^{1\mathrm{D}}$ is the observed linewidth and $c_\mathrm{s}$ the isothermal sound speed at 15~K. The Musca cloud is not covered by \citet{dame2001} and we adopt the Mach number for it from \citet{hacar2016musca}. We adopt 5~K as the 3-$\sigma$ uncertainty in the temperature, resulting in 22\% uncertainty (3-$\sigma$) in the sonic Mach number. In addition to the temperature uncertainty, it is unclear how well the Mach number derived from CO reflects the turbulent energy of the cloud. \cite{szucs2016} analysed numerical simulations that included gas-phase chemistry and radiative heating and cooling, and found that the observed CO line-widths are within 30--40\% of the true velocity dispersion. Taking this into account, we assign the total uncertainty of 60\% (3-$\sigma$) for the Mach number. 

% New data

To expand the sample provided by the above works, we additionally derive \rhopdf s for Orion A, Orion B, and California molecular clouds using the technique of \citet{kainulainen2014}. We summarise the derivation and present the resulting \rhopdf s in Appendix \ref{app:orion}. We adopt the CO linewidths for Orion A, Orion B, and California Cloud from \citet{dame2001}.  

In summary, our sample contains 15 Solar neighbourhood molecular clouds and includes most of the major cloud complexes within the distance of about 450~pc. The relevant parameters of the clouds are listed in Table~\ref{tab:parameters}.

\begin{table*}
\centering
\caption{Properties of molecular clouds}
\begin{tabular}{lcccccc}
\hline\hline
Cloud 		&\Ms   	&    \sigmas     	&   ref. 		& $b$ ($\beta = \infty$) 	&	 $b_\mathrm{MHD}$ ($\beta = 0.08$)\tablefootmark{a}	&	$b_\mathrm{MHD}$ ($\beta = 0.41$)\tablefootmark{b}	\\
			&		&       		& 			&					&	&		\\
\hline
California Cloud&      13	&	1.90		&	(1,1) 		& 0.46	&	1.68	&	0.85	\\
Cha I		&	7.1	&	1.76		&	(2,3) 		& 0.64	& 	2.37	&	1.20	\\ 
Cha II		& 	9.8	&	1.84		&	(2,3) 		& 0.54	& 	2.00	&	1.00	\\ 
Cha III		&	9.4	&	1.29		&	(2,3) 		& 0.22	& 	0.81	&	0.41	\\ 
CrA Cloud		&	5.6	&	2.08		&	(2,3) 		& 1.55	& 	5.70	&2.88	\\ 
LDN1228		&	9.0	&	1.85		&	(2,3) 		& 0.60 	&	2.22	&	1.12	\\ 
LDN1333		&	12	&	1.31		&	(2,3) 		& 0.18	&	0.66	&	0.34\\ 
LDN1719		&	6.9	&	1.52		&	(2,3) 		& 0.44	&	1.61	&	0.81\\
LDN204		&	7.9	&	1.61		&	(2,3) 		& 0.45	&	1.64	&	0.83\\	
Musca		&	4	&	1.32		&	(4,3)		& 0.54	& 	1.99	&	1.00	\\
Ophiuchus	&	8.2	&	1.81		&	(2,3) 		& 0.62	& 	2.26	&	1.14	\\	
Orion A		&	17	&	2.08		&	(1,1) 		& 0.50	&	1.84	&	0.93	\\	
Orion B		&	14	&	1.85		&	(1,1) 		& 0.39	&	1.44	&	0.72	\\	
Per Cloud		&	9.0	&	1.87		&	(2,3) 		& 0.63	& 	2.31	&1.16	\\	
Taurus		&	8.2	&	1.94		&	(2,3) 		& 0.79	& 	2.91	&	1.47	\\	
\hline
%\multicolumn{9}{l}{OMC-3 / MMS 7} \\   % From the top
\end{tabular}
\tablefoot{
\tablefoottext{a}{Using $B=8$ $\mu$G, $\rho = 100$ cm$^{-3}$, and $T=15$ K.}
\tablefoottext{b}{Using $B=5$ $\mu$G, $\rho = 200$ cm$^{-3}$, and $T=15$ K.}
}
\tablebib{(1)~This paper; (2) \citet{kainulainen_tan2013}; (3) \citet{kainulainen2014}; (4) \citet{hacar2016musca}.
}
\label{tab:parameters}
\end{table*}

%****************************************************
%****************************************************
\section{Results} %*
%****************************************************
%****************************************************
\label{sec:results}

% Describe the data: Ms - sigmas relationship

Figure~\ref{fig:b} presents the relationship between the sonic Mach numbers and density distribution widths in the 15 Solar neighbourhood clouds. The \Ms\ values span the range [4, 17] and the \sigmas\ values the range [1.29, 2.08]. It is not trivial to assess whether the variables show correlation, because the probability density functions of the uncertainties in the variables are not well known. The possibility of having a correlation can be probed by various tests. For example, ignoring all uncertainties, the Pearson's correlation coefficient for the data is 0.35 and the $p$-value 0.20, which indicates no correlation. Similarly, a linear fit using errors in both variables (the \textsf{fitexy} routine in IDL; \citealt{press1992}) results in a slope of $0.02 \pm 0.02$, indicating no significant correlation. However, we also cannot rule out a weak correlation (discussed further below). Clearly, including in the sample clouds over a wider range of Mach numbers, especially including extreme \Ms\ values, would be beneficial in further studies; we show in Appendix \ref{app:mc} that a sample of 50--100 clouds is sufficient to reliably establish a correlation, if present.  

% Comparison with predictions of Eq. (1).

We next consider the observed \Ms\ -- \sigmas\ data in the context of turbulence-regulated density variance. Figure~\ref{fig:b} shows the prediction given by Eq.~\ref{eq:b} for the hydrodynamic case ($\beta = \infty$) and for a moderately magnetized case ($\beta = 0.3$, i.e., $B \approx 4$~$\mu$G at $n=100$~cm$^{-3}$ and $T=15$~K). The comparison of these models with the observed data gives rise to two main results. 

First, the data points are in good general agreement with the ranges set by the models. The spread of the models that have no magnetic field ($b$ varies) cover all but one data point. If the magnetic field is stronger than given by $\beta \lesssim 0.05$ ($B \gtrsim 30$~$\mu$G at $n=1000$~cm$^{-3}$ and $T=15$~K), all but two clouds have \sigmas\ values in disagreement with the models. This implies a general agreement between the turbulence-regulated density structure and observations \emph{if the magnetic field is relatively weak}. 

Second, given the large uncertainties, the data cannot rule out a weak correlation such as the one predicted by Eq.~\ref{eq:b}; the correlation coefficient of the observed data is in agreement with a sample drawn from the model represented by Eq.~\ref{eq:b}  (demonstrated further in Appendix \ref{app:mc}). Further, the large uncertainties make it is possible to explain all the observations with a single \Ms\ -- \sigmas\ relationship (see Appendix \ref{app:mc}). A single \Ms -- \sigmas\ relationship corresponds to a family of parameters \{$b, \beta$\} (see Eq. \ref{eq:b}). This family can be characterized by a parameter $a = b^2 \beta /(1+\beta)$. We determined the best-fitting $a$ parameter by fitting Eq.~\ref{eq:b} to the observed data with $a$ as the free parameter. This yielded the best-fitting model with $a = 0.30 \pm 0.06$ (shown in Fig.~\ref{fig:b}, top panel, as the dotted line). The constraints the above results present for $b$ and $\beta$ are discussed in the following Section~\ref{sec:discussion}.

%****************************************************
%****************************************************
\section{Discussion} %*
%****************************************************
%****************************************************
\label{sec:discussion}

%  Interpretation of the observed variance.

Interpretation of the observed scatter in the \Ms -- \sigmas\ data remains uncertain because of the limitations of the data. All clouds can share one set of $\{\beta, b \}$ values, or they can have individually varying $\beta$ and/or $b$ values; the observations are in agreement with both cases. Disentangling between these two options would require more accurate measurements of \Ms\ and \sigmas\, or alternatively, a significantly larger sample of clouds. We next discuss how the $b$ and $\beta$ can be constrained in the above two cases.

\begin{figure}
\centering
\includegraphics[bb = 10 5 500 335, clip=true, width=0.42\textwidth]{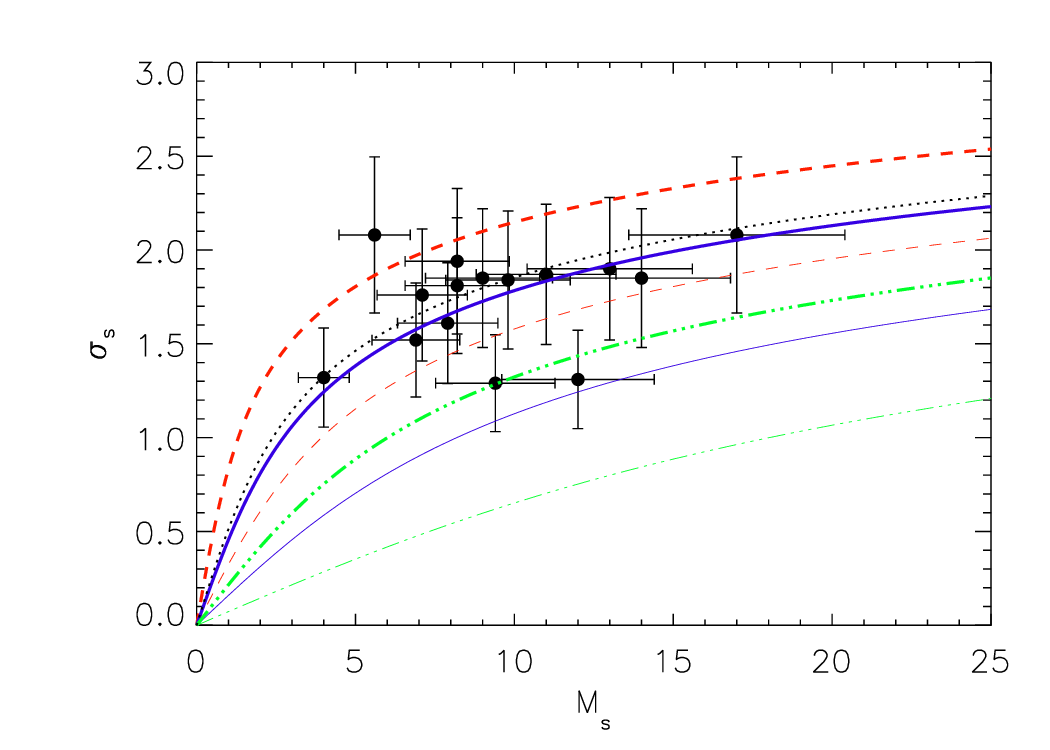}
\includegraphics[bb = 10 17 500 335, clip=true, width=0.42\textwidth]{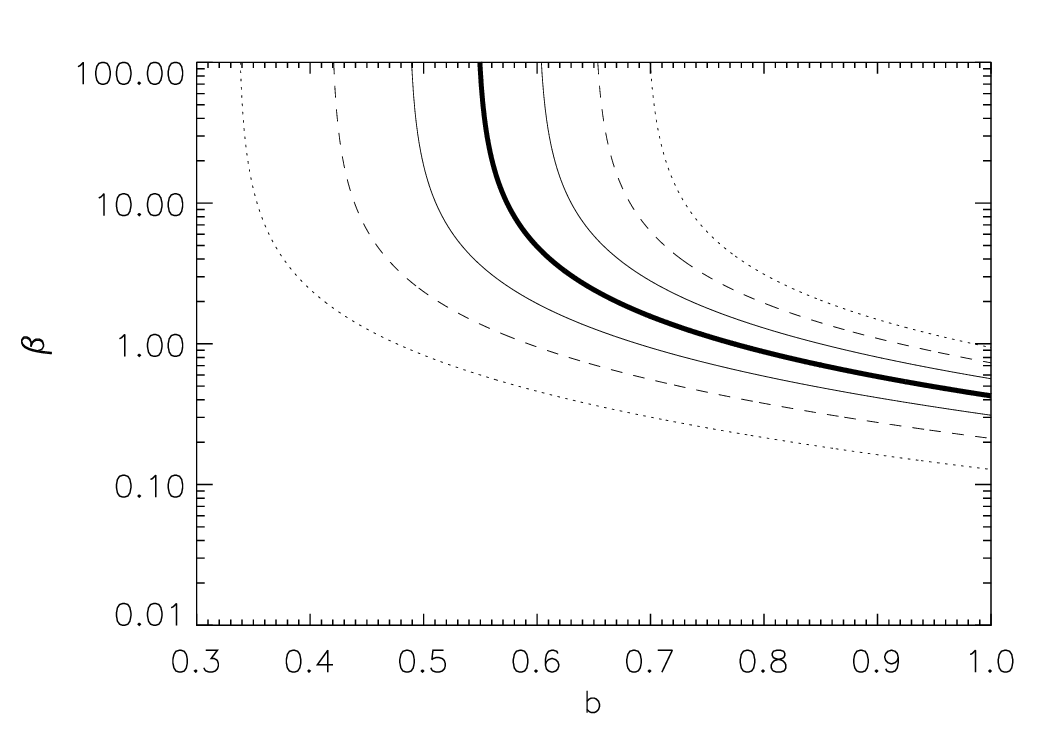}
\caption{\emph{Top: }Relationship between the density distribution width, \sigmas\, and sonic Mach number, \Ms. The red dashed curves show the case $\beta=\infty$, the blue curves $\beta = 0.3$, and green dash-dotted curves $\beta=$ 0.05. The thicker (and higher) curves correspond to $b=1$ and the thinner (and lower) curves to $b=1/3$. The dotted line shows the best-fitting model with $b^2\beta /(1+\beta) = 0.30 \pm 0.06$. \emph{Bottom: } The relationship between the magnetic field strength, $\beta$, and the compression parameter, $b$, for the model that best fits all data points. The thin solid, dashed, and dotted lines show the 1, 2, and 3-$\sigma$ confidence levels. 
              }
\label{fig:b}
\end{figure}

In the former case, i.e., if all clouds originate from one $\{\beta, b \}$ family, we can use the best-fitting model to constrain $\beta$ and $b$ (Fig.~\ref{fig:b}). The best-fitting model allows us to rule out several regimes: a regime of strong compression combined with a low magnetic field ($b \gtrsim$ 0.7 and $\beta \gtrsim$ 3), weak compression ($b \lesssim$ 0.35), or strong magnetic field ($\beta \lesssim 0.1$). 

If the $\{\beta, b \}$ values vary from cloud to cloud, we can obtain estimates for $b$ of each cloud by fixing $\beta$. Lower limits for $b$ result from the hydrodynamic case, i.e., $\beta = \infty$ (Table \ref{tab:parameters}). The lower limits are in agreement with the picture of turbulence-regulated density variance for 13 out of 15 clouds, i.e., they are between 0.3 and 1. The median lower limit for $b$ is 0.54.	

% 4: Contradiction with the Planck results???

Estimates of $b$ beyond the lower limits can be achieved if the magnetic field strengths are known. However, such measurements are very difficult to obtain. The most direct measurement is provided by the Zeeman-splitting measurements that indicate the maximum line-of-sight field strength of about 10 $\mu$G at densities lower than 1000~cm$^{-3}$ \citep{crutcher2012}. Alternatively, the mean magnetic field strengths for several nearby clouds have been estimated from the Planck data using the Davis-Chandrasekhar-Fermi technique \citep[CDF,][]{planck35}. The technique results in upper limits for the field strengths \citep[see the discussion in][]{planck35}. The field strengths reported in \citet{planck35} are 10--50~$\mu$G, in reasonable agreement with, but also systematically higher than, the results from Zeeman-splitting. These field strengths correspond to a lower-limit range for $\beta$ $\sim$0.002--0.05 in the density of 100 cm$^{-3}$ and temperature of 15 K. 

Given the uncertainty of the magnetic field strength estimates, we perform a simple exercise. We compute for each cloud a $b_\mathrm{MHD}$ value assuming the field strength of 8 $\mu$G (see Table \ref{tab:parameters}), broadly in agreement with Zeeman-splitting measurements and the upper limits from CDF estimates \citep{crutcher2012, planck35}. At the $\rho=100$ cm$^{-3}$ and $T=15$ K, $\beta = 0.08$ follows. Almost all $b_\mathrm{MHD}$ values exceed unity, which is in contradiction with the framework of Eq.~\ref{eq:b} \citep[values between $\sim$0.3--1 are expected,][]{federrath2008ks,federrath2010rk}. Low $b$ values are not in agreement with observations; 3 clouds out of 15 have $b$ $(\beta = \infty)$ below 0.4 and none have $b_\mathrm{MHD}$ below 0.4.

Several issues can contribute to the contradiction between the $b_\mathrm{MHD}$ values and Eq.~\ref{eq:b}. \citet{federrath2016brick} argues that it is the turbulent magnetic field component, not the total field strength, that is relevant for Eq.~\ref{eq:b}. This is because in the derivation of Eq.~(\ref{eq:b}), the magnetic pressure in the post-shock region enters \citep{molina2012}, which is likely dominated by the turbulent field component instead of the total field strength. The turbulent field component is often significantly smaller than the total field strength, especially in the presence of strong guide fields \citep{pillai2015,federrath2016brick,federrath2016jpp}. Using the turbulent field component (instead of the total field component) would bring down the $b_\mathrm{MHD}$ values, but the turbulent field component is difficult to measure observationally. Indeed, observational estimates of the magnetic field strength are usually very uncertain; it may be that our adopted values of 8 $\mu$G at 100 cm$^{-3}$ are not good estimates. Choosing instead 5 $\mu$G at 200 cm$^{-3}$ leads to $\beta = 0.41$ and to majority of the $b_\mathrm{mhd}$ values being smaller than or equal to unity (values given in Table \ref{tab:parameters}). It may also be that the density variance suffers from systematics; the employed technique was tested against numerical simulations \citep[see][]{kainulainen2014}, but those simulations do not necessarily capture all relevant aspects of real molecular clouds. Finally, the correspondence between \Ms\ and turbulence energy is also unclear, and could be further tested with numerical simulations \citep[e.g.,][]{szucs2016}.

%*********************************************************************************
%*********************************************************************************
 \section{Conclusions}
%*********************************************************************************
%*********************************************************************************

We analyzed the relationship between the \rhopdf\ width, \sigmas, and sonic Mach number, \Ms, in 15 Solar neighbourhood molecular clouds. We used data primarily from literature, but also derived new \rhopdf s for the Orion A, Orion B, and California molecular clouds. Our conclusions are as follows

\begin{enumerate}

\item The 15 clouds span the ranges \Ms\ = [4, 17] and \sigmas\ = [1.29, 2.08]. The variables show no significant correlation, however, because of the large scatter in the data we also cannot rule out a weak correlation. 

\item In the context of turbulence-dominated gas, the observed \Ms\ -- \sigmas\ relationship can constrain the turbulence driving parameter $b$ and the magnetic field strength measured by the thermal-to-magnetic pressure ratio, $\beta$. The majority of the clouds is not in agreement with field strengths stronger than $\beta < 0.05$, and none with $\beta < 0.01$. Considering the model that best fits the 15 clouds together, the data rule out the regime of strong compression combined with a weak magnetic field: $b \gtrsim$ 0.7 and $\beta \gtrsim$ 3. The data also rule out weak compression, $b \lesssim $ 0.35, or strong magnetic field, $\beta \lesssim 0.1$. 

\item When combined with magnetic field strength estimates, the \Ms\ -- \sigmas\ data rule out solenoidal driving with $b < 0.4$. However, most clouds have compression parameters larger than unity, which is not expected in the context of turbulence-regulated density variance (Eq.~\ref{eq:b}). It is not trivial to asses the source of the contradiction; one likely explanation is that it is not the total magnetic field strength, but the smaller-scale turbulent magnetic field component (i.e., the turbulent magnetic pressure in the post-shock regions) that affects the density variance based on the derivation of Eq.~\ref{eq:b} by \citet{padoan_nordlund2011} and \citet{molina2012}, as discussed in \citet[][]{federrath2016brick}. 

\end{enumerate}

Observational uncertainties restrict the determination of the parameters entering Eq. \ref{eq:b}, however, we demonstrate how systematic studies can start placing constraints on them; using our techniques, a sample of 50--100 clouds can establish the correlation between \sigmas\ and \Ms, if present (Appendix \ref{app:mc}).  

%---------------------------------------------------

\begin{acknowledgements}
This project has received funding from the European Union's Horizon 2020 research and innovation programme under grant agreement No~639459 (PROMISE). CF acknowledges funding provided by the Australian Research Council's Discovery Projects (grants DP150104329 and DP170100603) and by the ANU Futures Scheme.
\end{acknowledgements}

%---------------------------------------------------

	\bibliographystyle{aa} % style aa.bst
	\bibliography{ref_jk}

\begin{thebibliography}{34}
\expandafter\ifx\csname natexlab\endcsname\relax\def\natexlab#1{#1}\fi

\bibitem[{{Brunt} {et~al.}(2010){Brunt}, {Federrath}, \&
  {Price}}]{brunt2010fp_pdf}
{Brunt}, C.~M., {Federrath}, C., \& {Price}, D.~J. 2010, \mnras, 405, L56

\bibitem[{{Crutcher}(2012)}]{crutcher2012}
{Crutcher}, R.~M. 2012, \araa, 50, 29

\bibitem[{{Dame} {et~al.}(2001){Dame}, {Hartmann}, \& {Thaddeus}}]{dame2001}
{Dame}, T.~M., {Hartmann}, D., \& {Thaddeus}, P. 2001, \apj, 547, 792

\bibitem[{{Elmegreen} \& {Scalo}(2004)}]{elmegreen_scalo2004}
{Elmegreen}, B.~G. \& {Scalo}, J. 2004, \araa, 42, 211

\bibitem[{{Federrath}(2016)}]{federrath2016jpp}
{Federrath}, C. 2016, Journal of Plasma Physics, 82, 535820601

\bibitem[{{Federrath} \& {Banerjee}(2015)}]{federrath_banerjee2015}
{Federrath}, C. \& {Banerjee}, S. 2015, \mnras, 448, 3297

\bibitem[{{Federrath} \& {Klessen}(2012)}]{federrath_klessen2012}
{Federrath}, C. \& {Klessen}, R.~S. 2012, \apj, 761, 156

\bibitem[{{Federrath} {et~al.}(2008){Federrath}, {Klessen}, \&
  {Schmidt}}]{federrath2008ks}
{Federrath}, C., {Klessen}, R.~S., \& {Schmidt}, W. 2008, \apjl, 688, L79

\bibitem[{{Federrath} {et~al.}(2016){Federrath}, {Rathborne}, {Longmore},
  {Kruijssen}, {Bally}, {Contreras}, {Crocker}, {Garay}, {Jackson}, {Testi}, \&
  {Walsh}}]{federrath2016brick}
{Federrath}, C., {Rathborne}, J.~M., {Longmore}, S.~N., {et~al.} 2016, \apj,
  832, 143

\bibitem[{{Federrath} {et~al.}(2010){Federrath}, {Roman-Duval}, {Klessen},
  {Schmidt}, \& {Mac Low}}]{federrath2010rk}
{Federrath}, C., {Roman-Duval}, J., {Klessen}, R.~S., {Schmidt}, W., \& {Mac
  Low}, M.-M. 2010, \aap, 512, A81

\bibitem[{{Ginsburg} {et~al.}(2013){Ginsburg}, {Federrath}, \&
  {Darling}}]{ginsburg2013}
{Ginsburg}, A., {Federrath}, C., \& {Darling}, J. 2013, \apj, 779, 50

\bibitem[{{Goodman} {et~al.}(2009){Goodman}, {Pineda}, \&
  {Schnee}}]{goodman2009ps}
{Goodman}, A.~A., {Pineda}, J.~E., \& {Schnee}, S.~L. 2009, \apj, 692, 91

\bibitem[{{Hacar} {et~al.}(2016){Hacar}, {Kainulainen}, {Tafalla}, {Beuther},
  \& {Alves}}]{hacar2016musca}
{Hacar}, A., {Kainulainen}, J., {Tafalla}, M., {Beuther}, H., \& {Alves}, J.
  2016, \aap, 587, A97

\bibitem[{{Harvey} {et~al.}(2013){Harvey}, {Fallscheer}, {Ginsburg}, {Terebey},
  {Andr{\'e}}, {Bourke}, {Di Francesco}, {K{\"o}nyves}, {Matthews}, \&
  {Peterson}}]{harvey2013california}
{Harvey}, P.~M., {Fallscheer}, C., {Ginsburg}, A., {et~al.} 2013, \apj, 764,
  133

\bibitem[{{Hennebelle} \& {Chabrier}(2011)}]{hennebelle_chabrier2011}
{Hennebelle}, P. \& {Chabrier}, G. 2011, \apjl, 743, L29

\bibitem[{{Hennebelle} \& {Falgarone}(2012)}]{hennebelle_falgarone2012}
{Hennebelle}, P. \& {Falgarone}, E. 2012, \aapr, 20, 55

\bibitem[{{Jin} {et~al.}(2017){Jin}, {Salim}, {Federrath}, {Tasker}, {Habe}, \&
  {Kainulainen}}]{jin2017}
{Jin}, K., {Salim}, D.~M., {Federrath}, C., {et~al.} 2017, \mnras, 469, 383

\bibitem[{{Kainulainen} {et~al.}(2009){Kainulainen}, {Beuther}, {Henning}, \&
  {Plume}}]{kainulainen2009probing}
{Kainulainen}, J., {Beuther}, H., {Henning}, T., \& {Plume}, R. 2009, \aap,
  508, L35

\bibitem[{{Kainulainen} {et~al.}(2013){Kainulainen}, {Federrath}, \&
  {Henning}}]{kainulainen2013fh}
{Kainulainen}, J., {Federrath}, C., \& {Henning}, T. 2013, \aap, 553, L8

\bibitem[{{Kainulainen} {et~al.}(2014){Kainulainen}, {Federrath}, \&
  {Henning}}]{kainulainen2014}
{Kainulainen}, J., {Federrath}, C., \& {Henning}, T. 2014, Science, 344, 183

\bibitem[{{Kainulainen} \& {Tan}(2013)}]{kainulainen_tan2013}
{Kainulainen}, J. \& {Tan}, J.~C. 2013, \aap, 549, A53

\bibitem[{{K{\"o}rtgen} {et~al.}(2017){K{\"o}rtgen}, {Federrath}, \&
  {Banerjee}}]{koertgen2017driving}
{K{\"o}rtgen}, B., {Federrath}, C., \& {Banerjee}, R. 2017, \mnras, 472, 2496

\bibitem[{{Krumholz} \& {McKee}(2005)}]{krumholz_mckee2005}
{Krumholz}, M.~R. \& {McKee}, C.~F. 2005, \apj, 630, 250

\bibitem[{{Molina} {et~al.}(2012){Molina}, {Glover}, {Federrath}, \&
  {Klessen}}]{molina2012}
{Molina}, F.~Z., {Glover}, S.~C.~O., {Federrath}, C., \& {Klessen}, R.~S. 2012,
  \mnras, 423, 2680

\bibitem[{{Padoan} {et~al.}(2014){Padoan}, {Federrath}, {Chabrier}, {Evans},
  {Johnstone}, {J{\o}rgensen}, {McKee}, \& {Nordlund}}]{padoan2014ppvi}
{Padoan}, P., {Federrath}, C., {Chabrier}, G., {et~al.} 2014, Protostars and
  Planets VI, 77

\bibitem[{{Padoan} \& {Nordlund}(2011)}]{padoan_nordlund2011}
{Padoan}, P. \& {Nordlund}, {\AA}. 2011, \apj, 730, 40

\bibitem[{{Padoan} {et~al.}(1997){Padoan}, {Nordlund}, \&
  {Jones}}]{padoan1997nj}
{Padoan}, P., {Nordlund}, A., \& {Jones}, B.~J.~T. 1997, \mnras, 288, 145

\bibitem[{{Passot} \& {V{\'a}zquez-Semadeni}(1998)}]{passot1998}
{Passot}, T. \& {V{\'a}zquez-Semadeni}, E. 1998, \pre, 58, 4501

\bibitem[{{Pillai} {et~al.}(2015){Pillai}, {Kauffmann}, {Tan}, {Goldsmith},
  {Carey}, \& {Menten}}]{pillai2015}
{Pillai}, T., {Kauffmann}, J., {Tan}, J.~C., {et~al.} 2015, \apj, 799, 74

\bibitem[{{Planck Collaboration} {et~al.}(2016){Planck Collaboration}, {Ade},
  {Aghanim}, {Alves}, {Arnaud}, {Arzoumanian}, {Ashdown}, {Aumont},
  {Baccigalupi}, {Banday}, {Barreiro}, {Bartolo}, {Battaner}, {Benabed},
  {Beno{\^i}t}, {Benoit-L{\'e}vy}, {Bernard}, {Bersanelli}, {Bielewicz},
  {Bock}, {Bonavera}, {Bond}, {Borrill}, {Bouchet}, {Boulanger}, {Bracco},
  {Burigana}, {Calabrese}, {Cardoso}, {Catalano}, {Chiang}, {Christensen},
  {Colombo}, {Combet}, {Couchot}, {Crill}, {Curto}, {Cuttaia}, {Danese},
  {Davies}, {Davis}, {de Bernardis}, {de Rosa}, {de Zotti}, {Delabrouille},
  {Dickinson}, {Diego}, {Dole}, {Donzelli}, {Dor{\'e}}, {Douspis}, {Ducout},
  {Dupac}, {Efstathiou}, {Elsner}, {En{\ss}lin}, {Eriksen}, {Falceta-Gon{\c
  c}alves}, {Falgarone}, {Ferri{\`e}re}, {Finelli}, {Forni}, {Frailis},
  {Fraisse}, {Franceschi}, {Frejsel}, {Galeotta}, {Galli}, {Ganga}, {Ghosh},
  {Giard}, {Gjerl{\o}w}, {Gonz{\'a}lez-Nuevo}, {G{\'o}rski}, {Gregorio},
  {Gruppuso}, {Gudmundsson}, {Guillet}, {Harrison}, {Helou}, {Hennebelle},
  {Henrot-Versill{\'e}}, {Hern{\'a}ndez-Monteagudo}, {Herranz}, {Hildebrandt},
  {Hivon}, {Holmes}, {Hornstrup}, {Huffenberger}, {Hurier}, {Jaffe}, {Jaffe},
  {Jones}, {Juvela}, {Keih{\"a}nen}, {Keskitalo}, {Kisner}, {Knoche}, {Kunz},
  {Kurki-Suonio}, {Lagache}, {Lamarre}, {Lasenby}, {Lattanzi}, {Lawrence},
  {Leonardi}, {Levrier}, {Liguori}, {Lilje}, {Linden-V{\o}rnle},
  {L{\'o}pez-Caniego}, {Lubin}, {Mac{\'{\i}}as-P{\'e}rez}, {Maino},
  {Mandolesi}, {Mangilli}, {Maris}, {Martin}, {Mart{\'{\i}}nez-Gonz{\'a}lez},
  {Masi}, {Matarrese}, {Melchiorri}, {Mendes}, {Mennella}, {Migliaccio},
  {Miville-Desch{\^e}nes}, {Moneti}, {Montier}, {Morgante}, {Mortlock},
  {Munshi}, {Murphy}, {Naselsky}, {Nati}, {Netterfield}, {Noviello}, {Novikov},
  {Novikov}, {Oppermann}, {Oxborrow}, {Pagano}, {Pajot}, {Paladini},
  {Paoletti}, {Pasian}, {Perotto}, {Pettorino}, {Piacentini}, {Piat},
  {Pierpaoli}, {Pietrobon}, {Plaszczynski}, {Pointecouteau}, {Polenta},
  {Ponthieu}, {Pratt}, {Prunet}, {Puget}, {Rachen}, {Reinecke}, {Remazeilles},
  {Renault}, {Renzi}, {Ristorcelli}, {Rocha}, {Rossetti}, {Roudier},
  {Rubi{\~n}o-Mart{\'{\i}}n}, {Rusholme}, {Sandri}, {Santos}, {Savelainen},
  {Savini}, {Scott}, {Soler}, {Stolyarov}, {Sudiwala}, {Sutton}, {Suur-Uski},
  {Sygnet}, {Tauber}, {Terenzi}, {Toffolatti}, {Tomasi}, {Tristram}, {Tucci},
  {Umana}, {Valenziano}, {Valiviita}, {Van Tent}, {Vielva}, {Villa}, {Wade},
  {Wandelt}, {Wehus}, {Ysard}, {Yvon}, \& {Zonca}}]{planck35}
{Planck Collaboration}, {Ade}, P.~A.~R., {Aghanim}, N., {et~al.} 2016, \aap,
  586, A138

\bibitem[{{Press} \& {Teukolsky}(1992)}]{press1992}
{Press}, W.~H. \& {Teukolsky}, S.~A. 1992, Computers in Physics, 6, 274

\bibitem[{{Stutz} \& {Kainulainen}(2015)}]{stutz_kainulainen2015}
{Stutz}, A.~M. \& {Kainulainen}, J. 2015, \aap, 577, L6

\bibitem[{{Sz{\H u}cs} {et~al.}(2016){Sz{\H u}cs}, {Glover}, \&
  {Klessen}}]{szucs2016}
{Sz{\H u}cs}, L., {Glover}, S.~C.~O., \& {Klessen}, R.~S. 2016, \mnras, 460, 82

\bibitem[{{Vazquez-Semadeni}(1994)}]{vazquez1994}
{Vazquez-Semadeni}, E. 1994, \apj, 423, 681

\end{thebibliography}

%*****************************************************************
%*****************************************************************

\appendix

%*****************************************************************
%*****************************************************************
\section{\rhopdf s of Orion A, Orion B, and California Cloud}
%*****************************************************************
%*****************************************************************
\label{app:orion}

\begin{figure}
\centering
\includegraphics[width=\columnwidth]{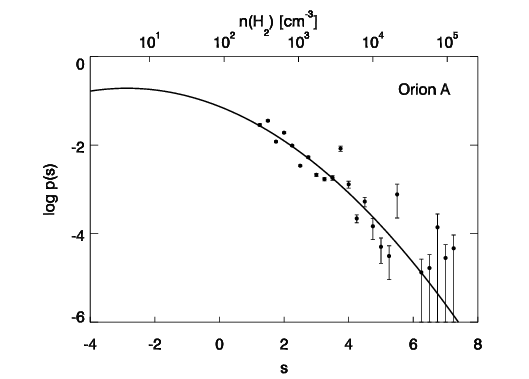}
\includegraphics[width=\columnwidth]{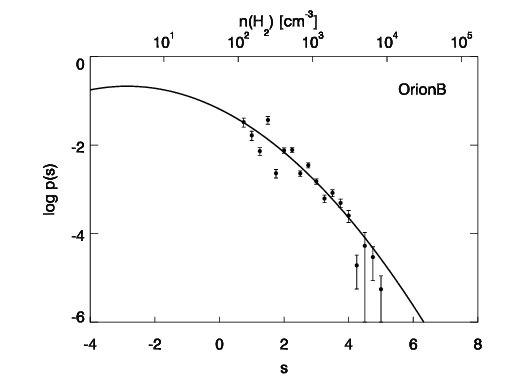}
\includegraphics[width=\columnwidth]{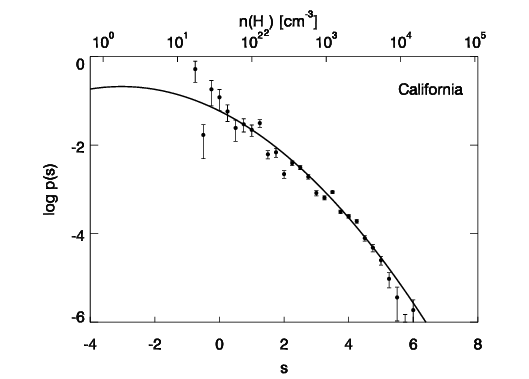}
\caption{\rhopdf s derived for Orion A (\emph{top}), Orion B (\emph{middle}), and California molecular clouds (\emph{bottom}). The solid line is a fit of a log-normal function to the data.}
\label{fig:orion}
\end{figure}

We derive \rhopdf s for the Orion A, Orion B, and California Clouds using the three-dimensional modelling technique presented by \citet{kainulainen2014} (see the paper for the details of the technique). As the data for the technique, we use for Orion A and the California nebula \emph{Herschel} dust emission derived column density data from \citet{stutz_kainulainen2015} and \citet{harvey2013california}, respectively. These data have a spatial resolution of full width half maximum ($FWHM$) = 38\arcsec, which corresponds to about 0.08~pc at the distance of Orion and the California nebula. This equals the physical resolution of the data used to derive the other \rhopdf s we employ in this work (see \cite{kainulainen2009probing}). For Orion B, we use a near-infrared extinction map from \citet{kainulainen2009probing} that has a resolution of \emph{FWHM}=150\arcsec. Acknowledging the caveat that this is coarser than the data for other clouds, we include the cloud in the sample. We tested the possible effect of resolution by smoothing the Orion A data with a factor of three and redoing the \rhopdf . This did not change the resulting \rhopdf\ width, although, it did truncate the \rhopdf\ as the highest column densities were smoothed out. The final \rhopdf s or Orion A, Orion B, and the California nebula are shown in Fig.~\ref{fig:orion} and the resulting \rhopdf\ widths are listed in Table~\ref{tab:parameters}.

%*****************************************************************
%*****************************************************************
\section{On the scatter of the observed \Ms\ -- \sigmas\ relationship}
%*****************************************************************
%*****************************************************************
\label{app:mc}

We analyze here the scatter in the observed \Ms\ -- \sigmas\ relationship to address three questions: 1) whether or not the data can be well fitted by a single model represented by Eq.~\ref{eq:b}; 2) can the observations rule out a correlation between \Ms\ and \sigmas\ such as the one predicted by Eq.~\ref{eq:b}, and 3) given the uncertainties, can a larger sample reliably establish the presence of a relationship described by Eq.~\ref{eq:b}, if present.

We address the first two questions with a Monte Carlo simulation in which we draw random samples of 15 data points from the one model (Eq.~\ref{eq:b}) that best fits all the observed data points (see Fig.~\ref{fig:b}). The model is characterized by the parameter $a = 0.30 \pm 0.06$. We restrict the draw between \Ms\ = 4--20. We draw 15 $\{$\Ms, \sigmas $\}$ pairs from the model so that the probability $P$ along the model curve, i.e., $dP / df$ where $f = (\ln{1+b^2 {\cal M}_\mathrm{s}^2 \beta / (1+\beta)})^{1/2}$, is constant and uniform. The data points are then assigned an error, according to the observational uncertainties, i.e., 20\% for both \sigmas\ and \Ms. This procedure is repeated 10$^4$ times to obtain reasonable statistics. 

\begin{figure}
\centering
\includegraphics[width=\columnwidth]{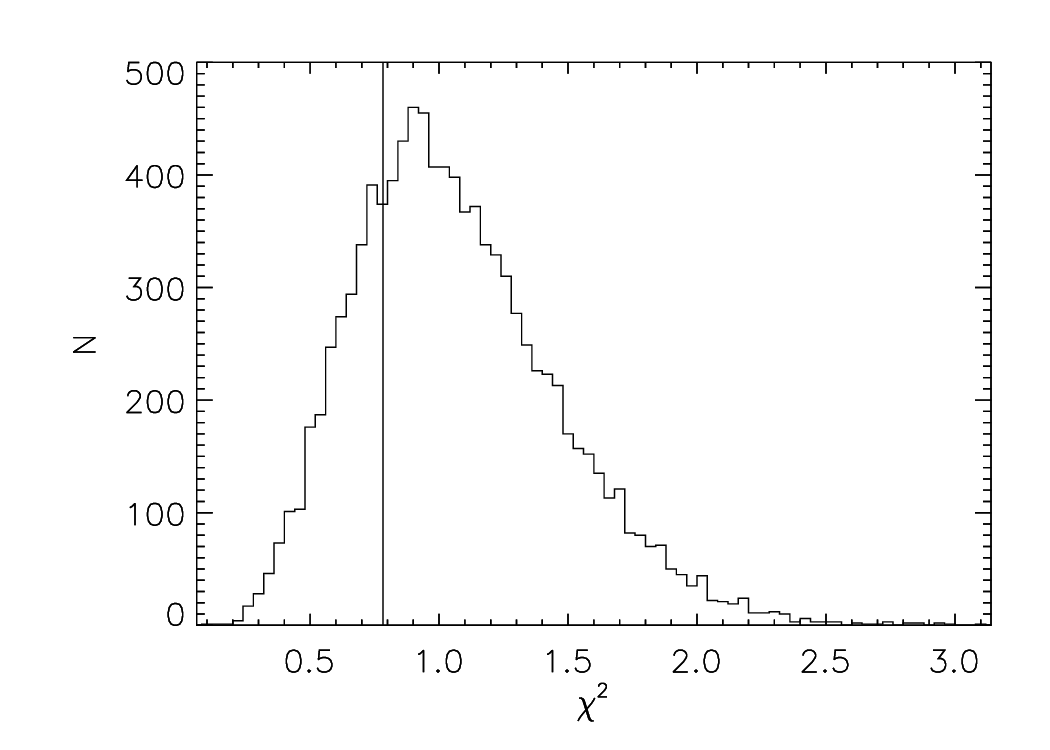}
\includegraphics[width=\columnwidth]{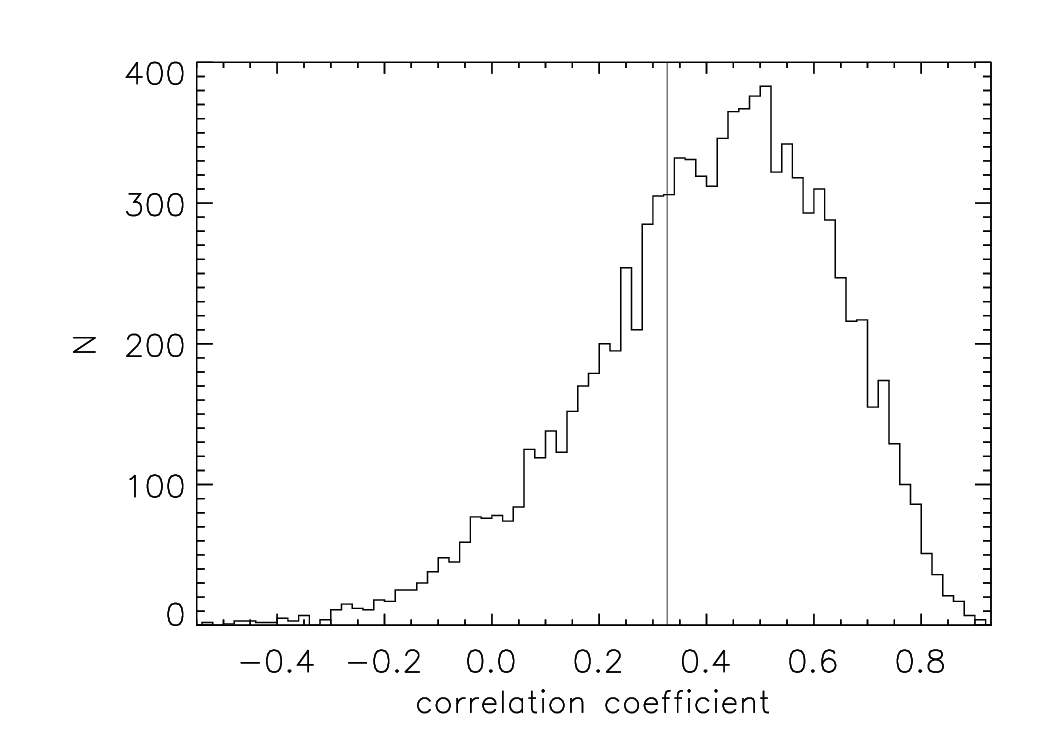}
\includegraphics[width=\columnwidth]{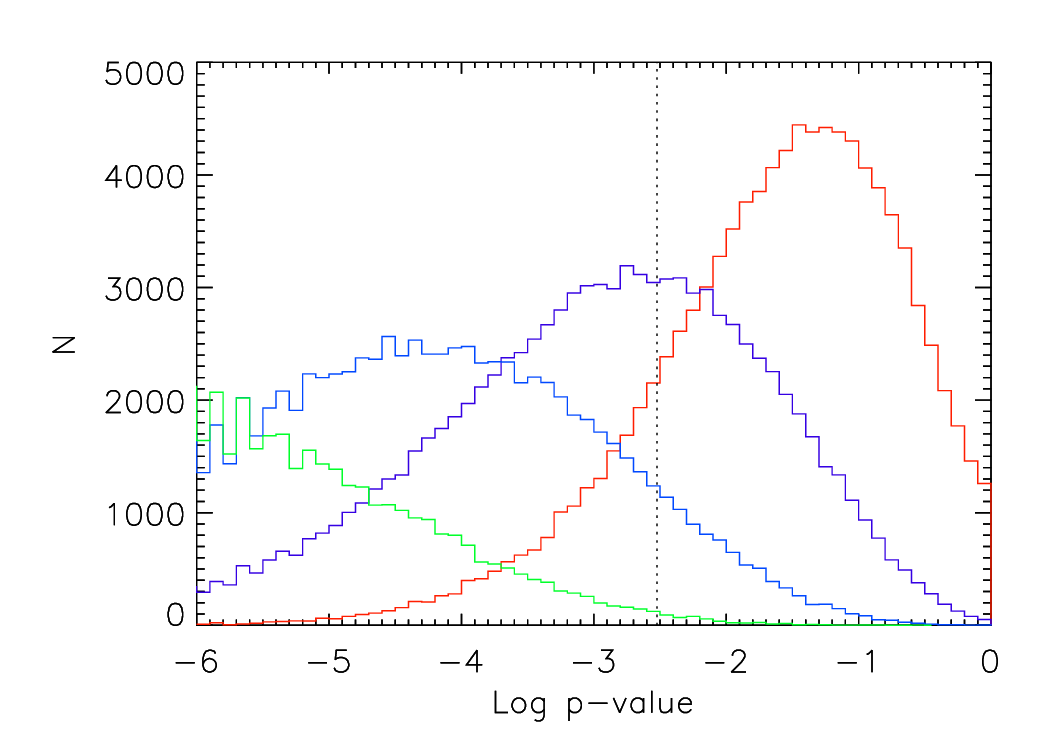}
\caption{\emph{Top: } The distribution of the reduced chi-square values in the simulation. The solid vertical lines show the values from the observed data. \emph{Middle: }The distribution of the Pearson's correlation coefficients in the simulation. The limit for significant correlation is about 0.7. The solid vertical lines show the values from the observed data. \emph{Bottom: }Distributions of $p$-values in a simulation with 15 (red), 30 (purple), 50 (blue), and 80 (green) clouds. The dotted line shows the $p$-value 0.003.}
\label{fig:mc}
\end{figure}

For each repetition, we computed the reduced chi-square value between the model and the simulated data points. The distribution of chi-square is shown in Fig.~\ref{fig:mc}, together with the chi-square of the observations and the best-fitting model. It shows that the observed chi-square does not strongly deviate from the distribution of the chi-square of the simulation. This indicates that the observed \Ms\ -- \sigmas\ relationship \jk{can} be explained with a single model represented by Eq.~\ref{eq:b}.

For each repetition, we also computed the Pearson's correlation coefficient. The resulting distribution of  coefficients is shown in Fig.~\ref{fig:mc}. For reference, a correlation coefficient higher than $\sim$0.7 is usually taken to indicate a significant correlation ($p < 0.003$). The observed correlation coefficient is well within the range of the simulated coefficients. Importantly, the distribution shows that in $\sim$8\% of the cases a significant correlation is detected. This indicates that we cannot rule out a correlation such as predicted by Eq.~\ref{eq:b} with our data, and in fact, with only 15 data points we would not expect to be able to do so (given the uncertainties of the measurements).

Finally, we address the third question with the following exemplary setup. An observational sample that has the best chance of detecting a correlation between \sigmas\ and \Ms, if present, is one that probes extreme Mach numbers, both small and large. With that in mind, consider a sample of $N_\mathrm{clouds}$, half of which are chosen between \Ms$=[5, 10]$ and the other half between $[15, 20]$. How large a sample is needed to detect a correlation, if any, assuming the uncertainties of our current observational sample? To find this out, we repeated the Monte Carlo simulation described above, only by changing the number of data points, $N_\mathrm{clouds}$, between 15 and 200, and restricting the Mach numbers at the ranges 5--10 and 15--20. For each simulation characterised by $N_\mathrm{clouds}$, we saved the correlation coefficient and computed the corresponding $p$-value. Figure \ref{fig:mc} shows as an example histograms for $N_\mathrm{clouds}=15, 30, 50$, and 80. The experiment shows that as the sample size grows, the ability to detect the correlation of the variables improves. With 80 clouds, the fraction of $p$-values above 0.003 is less than 0.5\%. Even if this experiment is simplistic, it indicates that using our adopted techniques, a well-constructed sample of some 50--100 clouds will be sufficient to detect the correlation between \sigmas\ and \Ms, if present.

\end{document}